# YBCO Ceramic Nanofibers obtained by the new technique of Solution Blow Spinning

**M. Rotta, L. Zadorosny, C. L. Carvalho, J. A. Malmonge, L. F. Malmonge, R. Zadorosny**

Departamento de Física e Química, Univ Estadual Paulista-UNESP, Caixa Postal 31, 15385-000, Ilha Solteira, SP, Brazil

**ABSTRACT**

This study proposes a novel solution blow spinning technique (SBS) for fabricating YBCO ceramic nanofibers. The precursor solutions were obtained from Y, Ba, and Cu metallic acetates (Ac) and poly(vinyl pyrrolidone) (PVP, Mw = 360,000). Ac:PVP concentrations of 1:1 and 5:1 were tested, resulting in ceramic nanofibers with average diameters of 359 and 375 nm, respectively. X-ray diffraction confirmed the formation of a pure phase of $YBa_2Cu_3O_{7-x}$. This is the first study to use SBS for fabricating YBCO nanofibers, and this technique shows promise for obtaining high-quality ceramic materials.

**1. Introduction**

Novel techniques are urgently required for fabricating materials owing to developments in nanotechnology and miniaturized devices [1,2]. Currently, electrospinning (ES) is used for fabricating one-dimensional nanostructured materials such as wires [3], tubes [4], and fibers [5] by applying high voltage between a tip and a collector to disperse a polymeric solution and create nanofibers [6–14] . Recently, a simple technique called solution blow spinning (SBS) has attracted attention for producing one-dimensional polymeric and ceramic samples [15,16]. SBS can produce tens of times more nanofibers than ES with improved cost/benefit ratio [17,18]. It does not require high voltages or a conducting collector. Furthermore, it can be used in a variety of polymeric solutions, regardless of their dielectric constant and heat sensitivity, such as some proteins [17].

Oxide ceramics of the Y-Ba-Cu-O (YBCO) system are an interesting class of materials that has not yet been studied in the framework of SBS [6,19–21]. YBCO, owing to their superconducting properties, have been applied in nanoscale molecular circuits [22], high-frequency electronics [23,24], power transmission [25], transformers [26], generators [27], and motors [28].

This study produced high-quality YBCO nanofibers by SBS. Samples were successfully obtained with an average diameter of 368 nm, and X-ray diffraction (XRD) analysis confirmed the formation of a pure phase. To the knowledge of the authors, this is the first study to produce such a ceramic system by SBS.

The outline of this paper is as follows. First, we provide an overview of the materials and method used to obtain the precursor solution and, consequently, the nanofibers. Then, we describe the processes applied for the thermal treatment and characterization of the samples. The remainder of the paper presents and discusses the obtained results.

## 2. Experimental Procedures

### 2.1 Materials

Sol-gel precursor solutions were produced by using the following reagents: poly(vinyl pyrrolidone) (PVP, Mw = 360,000), yttrium acetate hydrate [$Y(CH_3CO_2)_3xH_2O$] (99.9%) from Sigma, and barium acetate [$Ba(CH_3CO_2)_2$] (99%) and copper acetate monohydrate [$Cu(CH_3CO_2)_2 H_2O$] (99%), both from Sigma Aldrich. Before the reagents were weighted, they were dried overnight at 100°C.

### 2.2 Sol-gel process

Initially, PVP was dissolved in a solution of 65% methanol, 21% acetic acid, and 14% propionic acid at a concentration of 5 wt%. Then, acetates were added to obtain the correct stoichiometric composition of Y:Ba:Cu = 1:2:3. The final solution, called the "precursor solution," was stirred at room temperature for 24 h in a hermetic vessel to obtain a stable and homogeneous solution for spinning. As expected, by keeping a fixed PVP concentration, e.g., 5 wt%, the morphologies and fiber diameters of the produced samples showed similar values [18], as discussed in Section 3. However, to confirm the formation of the composite and YBCO nanofibers (after the polymer burns out), we tested Ac:PVP precursor concentrations of 1:1 and 5:1. A detailed study of different acetate:poly(vinyl alcohol) (Ac:PVA) weight ratios has been conducted previously [19].

**2.3 Solution blow spinning**

The SBS technique uses two concentric nozzles: the inner one injects the polymeric solution at a controlled rate, and the outer one flows high-pressure gas for dragging the fibers to a rotating cylindrical collector. The angular speed and distance from the injection point to the collector (working distance) are set as 40 rpm and 40 cm, respectively. Figure 1 shows the SBS experimental apparatus.

The precursor solution was loaded in a 3-mL hypodermic needle, which in turn was injected in a needle with 0.5-mm internal diameter. To produce Ac:PVP samples with 1:1 and 5:1 concentration, injection rates of 50 and 60 µL/min with air pressures of 133 and 66 kPa, respectively, were used. A halogen lamp was used to increase the local temperature and to evaporate the solvents as the solution was dragged. The nanofibers in the green state (before heat-treatment) were collected in a coated cylinder covered by a steel screen.

**2.4 Heat treatment**

The nanofiber composite in the green state was dried overnight at 100°C and then heat-treated in two steps. In the first step, the samples were calcined at T = 450°C for 3 h in an oven to eliminate organic material. Both heating and cooling were performed at 1°C/min. In the second step, the samples were sintered in a tube furnace at 820°C for 14 h and then at 925°C for 1 h, following which they were treated at 725°C for 3 h and 450°C for 12 h. Both steps were performed under oxygen flow. The heating rates from room temperature to 820°C and from 820°C to 925°C were 3°C/min and 1°C/min, respectively. The cooling rate from 925° to 750°C was 1°C/min, and that from 750°C to room temperature (remaining 12 h at 450°C) was 3°C/min. The samples obtained with Ac:PVP concentrations of 1:1 and 5:1 are called S11 and S51, respectively.

**2.5 Characterization**

Scanning electron microscopy (SEM) analysis was performed using an EVO LS15 Zeiss operated at 20 kV. Green state samples were coated with a thin layer of gold before SEM observations. IMAGE J software was used to measure the diameters of 100 nanofibers. Thermogravimetric measurements were performed using a TA Instruments model Q600 at temperatures between 25 and 1000°C at a heating rate of 10°C/min. Analyses were performed in nitrogen atmosphere with flow rate of 100 mL/min. XRD was performed using a Shimadzu XDR-6000 diffractometer with CuK$_α$ radiation (wavelength: 1.5418 Å). The displacement ranged from 2θ = 5° to 60° at a scan rate of 0.02°/min.

**3. Results and discussion**

Fibers with a continuous structure are obtained when the polymer concentration is sufficient to thoroughly entangle the long polymeric chains with the molecules of the solvent [18]. In addition, a polymeric solution with low surface tension needs to be obtained to facilitate the evaporation of the solvent and then dry the solution before it has been collected [18]. Therefore, the morphology of the nanofibers is strongly influenced by the PVP concentration in the precursor solution. On the other hand, depending on the polymer concentration in the solution, the injection rate and pressure of compressed air need to be adjusted. Then, a stable cone formed by the solution at the tip can be established and, consequently, continuous fibers structures are produced. If suitable parameters are not set, undesired drops or discontinuities in the jet or even a film could be produced [18].

The diameter of the fibers is also influenced by the polymer concentration in the solution, i.e., lower polymer concentration produces fibers with smaller diameters. The evaporation rate of the solvent also depends on the polymer concentration, and consequently, beads and interconnections may be produced between the produced fibers [16-18].

After setting suitable parameters, polymeric nanofibers such as those shown in Figure 2 were produced. For as produced S51, the fibers were oriented randomly and presented several beads, as shown in the micrograph in Figure 2(a). The average diameter of these fibers was 389 nm, and around 25% of them had diameters smaller than 300 nm, as shown in Figure 2(b).

For as produced S11, the production of beads decreased, as shown in the micrograph in Figure 2(c). The average diameter of these fibers was 426 nm, as shown in Figure 2(d); this value was close to that of S51. However, only 7% of the nanofibers showed diameters smaller than 300 nm. This behavior was expected because we used the same polymer concentration in both solutions, which resulted in fibers with similar diameters [17,18].

The commercial PVP sample shows weight loss of 6% at ~100°C owing to the evaporation of water, as indicated by the TGA curve in Figure 3(a). The second event with an endothermic peak in the DTA curve at 420°C is attributed to PVP decomposition. From 25 to 280°C, samples S11 and S51 lost 15.5% and 22.4% of their initial mass, respectively, as shown in Figures 3(b) and (c). This behavior is attributed to dehydration and the elimination of other volatile substances [4]. These figures also show successive endothermic peaks from 280 to 500°C for both samples. These events correspond to polymer combustion and decomposition of the organic groups of the acetates used in the precursor solutions [4,6,21]. In the same temperature range, the S11 sample lost more mass (53%) than the S51 one (33%), which is related the larger amount of PVP in its composition. Therefore, the peaks associated with acetate degradation are masked in the thermogram of sample S11. The exothermic peak above 780°C (DTA curve) indicates the beginning of YBCO crystallization [6,21]. The total mass lost from room temperature up to 925°C was 94.2%, 74%, and 64% in the commercial PVP, S11, and S51 samples, respectively. These results were used to choose the heat-treatment temperatures applied to obtain the YBCO phase.

Figure 4 shows SEM images of the S51 and S11 samples and the distributions of their diameters after all heat treatments and oxygenation. The S51 and S11 samples showed a fibrous structure with nanoscale dimensions and average diameters of 359 and 375 nm, respectively. In the S11 sample, 32% of nanofibers were smaller than 300 nm; in the S51 sample, this value decreases to 27%.

The ceramic nanofibers are not smooth like those obtained by the composites shown in Figure 2. Instead, they seem to be made of grains that are joined to each other, as seen in the inset of Figure 4(a). Along with the samples in the green state, the ceramic nanofibers showed very similar diameters.

It should be noted that in our experiment, the green nanofibers were produced using flow rates of 50 and 60 µL/min for the S11 and S51 samples, respectively. These values are around four times greater than those used in the ES experiments [21].

XRD patterns of the green nanofiber and heat-treated samples S51 and S11 are shown in Figure 5. The green sample shows an amorphous pattern with a broad peak at $2\theta = 8°$ that is attributed to PVP [4,29]. The XRD patterns of samples S51 and S11 showed peaks that are characteristic of YBCO ceramic (JCPDS-78-2273, curve (d)). We did not observe the formation of secondary phases, indicating that the heat-treatment applied was adequate. The Miller index in Figure 5 is related to the patterns shown in Figures 5(b), (c), and (d).

## 4. Conclusions

Nanofibers of pure YBa$_2$Cu$_3$O$_7$ were successfully obtained, for the first time, by using the novel SBS technique. Polymeric solutions with Ac:PVP concentrations of 5:1 and 1:1 were prepared. After heat-treatment, YBCO ceramic nanofibers with average diameters of 359 and 375 nm were produced from the 5:1 and 1:1 samples, respectively. We showed that the SBS technique can produce YBCO ceramic nanofibers with a production rate close to four times greater that of ES.

**References**

[1] C.A. Gautier, J.C. Loulergue, J. Etchepare, S. Hunsche, H. Kurz, T.P. Doughtery, S. Adachi, K.A. Nelson, Y.R. Shen, T. Ruf, J.M. Zhang, R. Lauck, M. Cardona, T. Hattori, Y. Homma, A. Mitsuishi, M. Tacke, A. Torabi, T.E. Stevens, M. Born, K. Huang, Directed Assembly of One-Dimensional Nanostructures into Functional Networks, Science. 291 (2001) 630–634.

[2] G.Y. Tseng, J.C. Ellenbogen, Toward Nanocomputers, Science, 294 (2001) 1293–1295.

[3] E.A. Duarte, P.A. Quintero, M.W. Meisel, J.C. Nino, Electrospinning synthesis of superconducting BSCCO nanowires, Phys. C Supercond. Its Appl. 495 (2013) 109–113.

[4] Z. Shen, Y. Wang, W. Chen, L. Fei, K. Li, H.L.W. Chan, L. Bing, Electrospinning preparation and high-temperature superconductivity of YBa2Cu3O7-x nanotubes, J. Mater. Sci.; 48 (2013) 3985–3990.

[5] J. Yuh, L. Perez, W.M. Sigmund, J.C. Nino, Electrospinning of complex oxide nanofibers, Phys. E Low-Dimensional Syst. Nanostructures; 37 (2007) 254–259.

[6] J. Yuh, L. Perez, W.M. Sigmund, J.C. Nino, Sol-gel based synthesis of complex oxide nanofibers, J. Sol-Gel Sci. Technol. 42 (2007) 323–329.

[7] A. Rogina, Electrospinning process: Versatile preparation method for biodegradable and natural polymers and biocomposite systems applied in tissue engineering and drug delivery, Appl. Surf. Sci. 296 (2014) 221–230.

[8] C. Mu, Y. Song, X. Wang, P. Wu, Kesterite Cu2ZnSnS4 compounds via electrospinning: A facile route to mesoporous fibers and dense films, J. Alloys Compd. 645 (2015) 429–435.

[9] G.C. Leindecker, A.K. Alves, C.P. Bergmann, Synthesis of niobium oxide fi bers by electrospinning and characterization of their morphology and optical properties, Ceram. Int.

(2014) 1–6.

[10]   N. Bhardwaj, S.C. Kundu, Electrospinning : A fascinating fi ber fabrication technique, Biotechnology Advances; 28 (2010) 325–347.

[11]   R. Nayak, R. Padhye, I.L. Kyratzis, Y.B. Truong, L. Arnold, Recent advances in nanofibre fabrication techniques, Textile Research Journal. (2011) 82(2):129-47.

[12]   Z. Huang, Y. Zhang, M. Kotaki, S. Ramakrishna, A review on polymer nanofibers by electrospinning and their applications in nanocomposites, Composites Science and Technology; 63 (2003) 2223–2253.

[13]   J. Gao, W. Li, H. Shi, M. Hu, R.K.Y. Li, Preparation , morphology , and mechanical properties of carbon nanotube anchored polymer nanofiber composite, Compos. Sci. Technol. 92 (2014) 95–102.

[14]   W. Teo, S. Ramakrishna, Electrospun nanofibers as a platform for multifunctional , hierarchically organized nanocomposite, Compos. Sci. Technol. 69 (2009) 1804–1817.

[15]   R. Maria, R. Rodrigues, J. Elvis, Production of submicrometric fi bers of mullite by solution blow spinning ( SBS ), Mater. Lett. 149 (2015) 47–49.

[16]   B. Cheng, X. Tao, L. Shi, G. Yan, X. Zhuang, Fabrication of ZrO 2 ceramic fi ber mats by solution blowing process, Ceram. Int. 40 (2014) 15013–15018.

[17]   E.S. Medeiros, G.M. Glenn, A.P. Klamczynski, W.J. Orts, L.H.C. Mattoso, Solution Blow Spinning : A New Method to Produce Micro- and Nanofibers from Polymer Solutions, Journal of Applied Polymer Science. (2009) 113(4):2322-30.

[18]   S. Afonso, J.E. Oliveira, E.A. Moraes, R.G.F. Costa, L.H.C. Mattoso, W.J. Orts, E.S. Medeiros, W. Luis, C. Ufscar, R.W. Luis, Nano and Submicrometric Fibers of Poly ( D , L -Lactide ) Obtained by Solution Blow Spinning : Process and Solution Variables, Journal of Applied Polymer Science. (2011) 122(5):3396-405.

[19]   X.M. Cui, W.S. Lyoo, W.K. Son, D.H. Park, J.H. Choy, T.S. Lee, W.H. Park, Fabrication of YBa 2 Cu 3 O 7−δ superconducting nanofibres by electrospinning, Supercond. Sci. Technol. Supercond. Sci. Technol. 19 (2006) 1264–1268.

[20]   E. Zussman, YBCO nanofibers synthesized by electrospinning a solution of poly ( acrylic acid ) and metal nitrates, (2008) 1664–1668.

[21]   E.A. Duarte, N.G. Rudawski, P.A. Quintero, M.W. Meisel, J.C. Nino, Electrospinning of superconducting YBCO nanowires, Superconductor Science And Technology, (2014);

28(1):015006-12.

[22] R.F. Service, N. York, S. Rf, P. Types, C.S. Mostrando, Nanocomputing. Assembling nanocircuits from the bottom up. Science. 293 (2016) 1–2.

[23] S. Anders, M.G. Blamire, F. Buchholz, D. Crété, R. Cristiano, P. Febvre, L. Fritzsch, A. Herr, E. Il, J. Kohlmann, J. Kunert, H. Meyer, J. Niemeyer, T. Ortlepp, H. Rogalla, T. Schurig, M. Siegel, R. Stolz, E. Tarte, H.J.M. Brake, H. Toepfer, J. Villegier, A.M. Zagoskin, A.B. Zorin, European roadmap on superconductive electronics – status and perspectives q, Phys. C Supercond. Its Appl. 470 (2010) 2079–2126.

[24] P. Febvre, G. Burnell, T. Claeson, D. Cre, G.J. Gerritsma, H. Hilgenkamp, R. Humphreys, Z. Ivanov, W. Jutzi, M.I. Khabipov, J. Mannhart, J. Niemeyer, A. Ravex, H. Rogalla, M. Russo, J. Satchell, M. Siegel, H. To, F.H. Uhlmann, E. Wikborg, D. Winkler, A.B. Zorin, SCENET roadmap for superconductor digital electronics, 439 (2006) 1–41.

[25] T. Izumi, Y. Shiohara, R & D of coated conductors for applications in Japan, Phys. C Supercond. Its Appl. 470 (2010) 967–970.

[26] H. Zueger, 630 kVA high temperature superconducting transformer, Cryogenics. 38 (1998) 1169–1172.

[27] P.N. Barnes, M.D. Sumption, G.L. Rhoads, Review of high power density superconducting generators : Present state and prospects for incorporating YBCO windings, Cryogenics. 45 (2005) 670–686.

[28] P. Manuel, Prospects for application of high temperature superconductors to electric power networks, Physica C: Superconductivity. 376 (2002) 1591–1597.

[29] D. Khan, B. Ahmed, S.K. Raghuvanshi, M.A. Wahab, Structural , morphological and optical properties of silver doped polyvinylpyrrolidone composites, Indian Journal of Pure & Applied Physics. 52 (2014) 192–197.

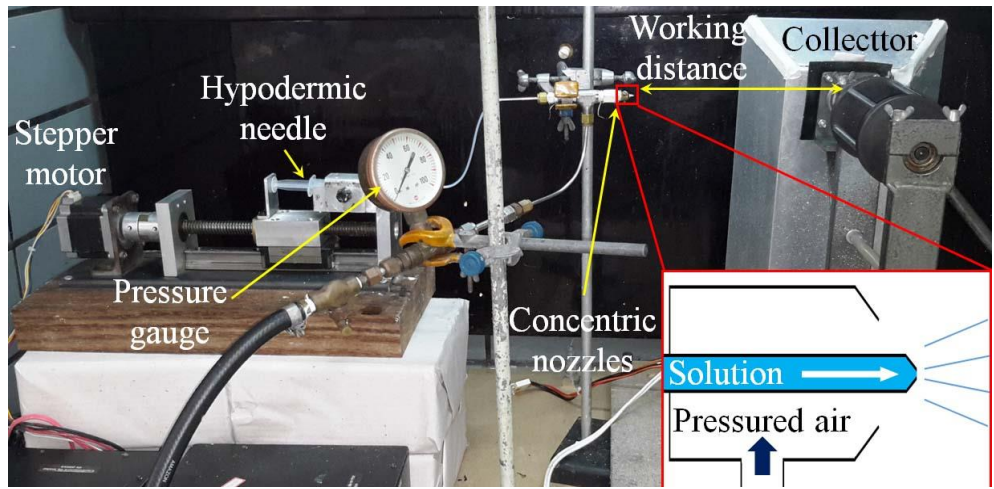

Figure 1: SBS experimental apparatus. The stepper motor pushes the hypodermic needle loaded with the precursor solution. Then, the pressured air in the outer nozzle drags the solution in the inner nozzle. As the solution travels along the work distance, the solvents evaporate and nanofibers are formed. In the rotating collector, a thin blanket with nanofibers is obtained. For more details, see Ref. [17,18].

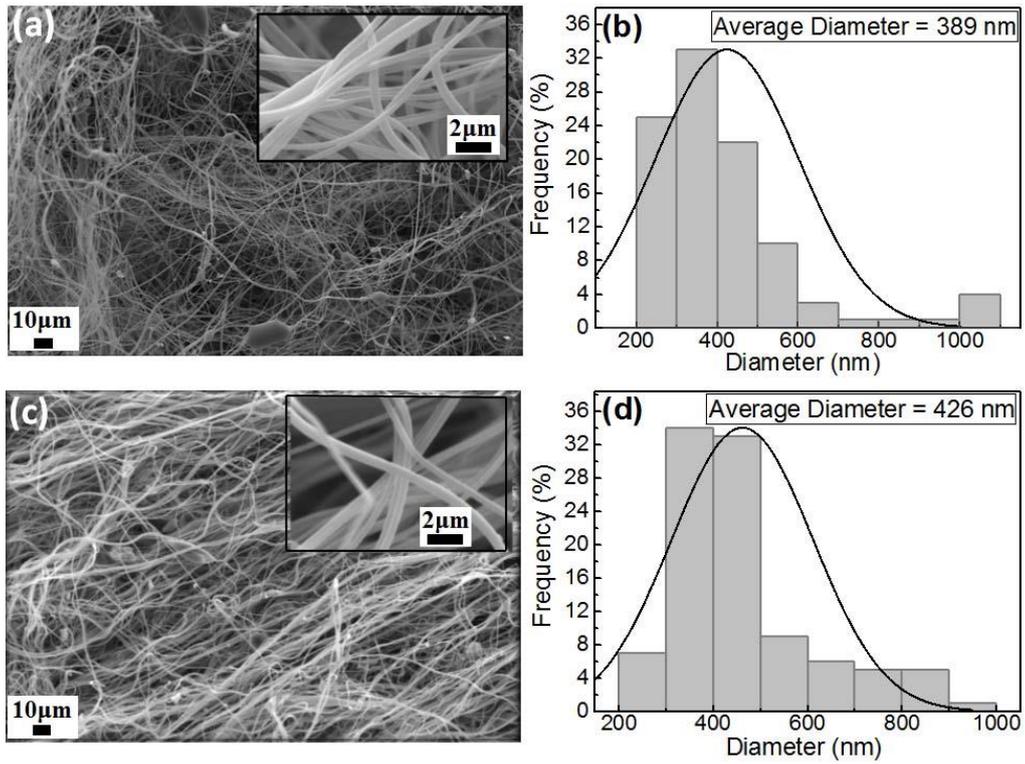

Figure 2: (a) SEM micrographs of the as produced S51 sample and (b) the size distribution of the S51 nanofibers. (c) SEM micrographs of the as produced S11 sample and (d) the size distribution of the S11 nanofibers.

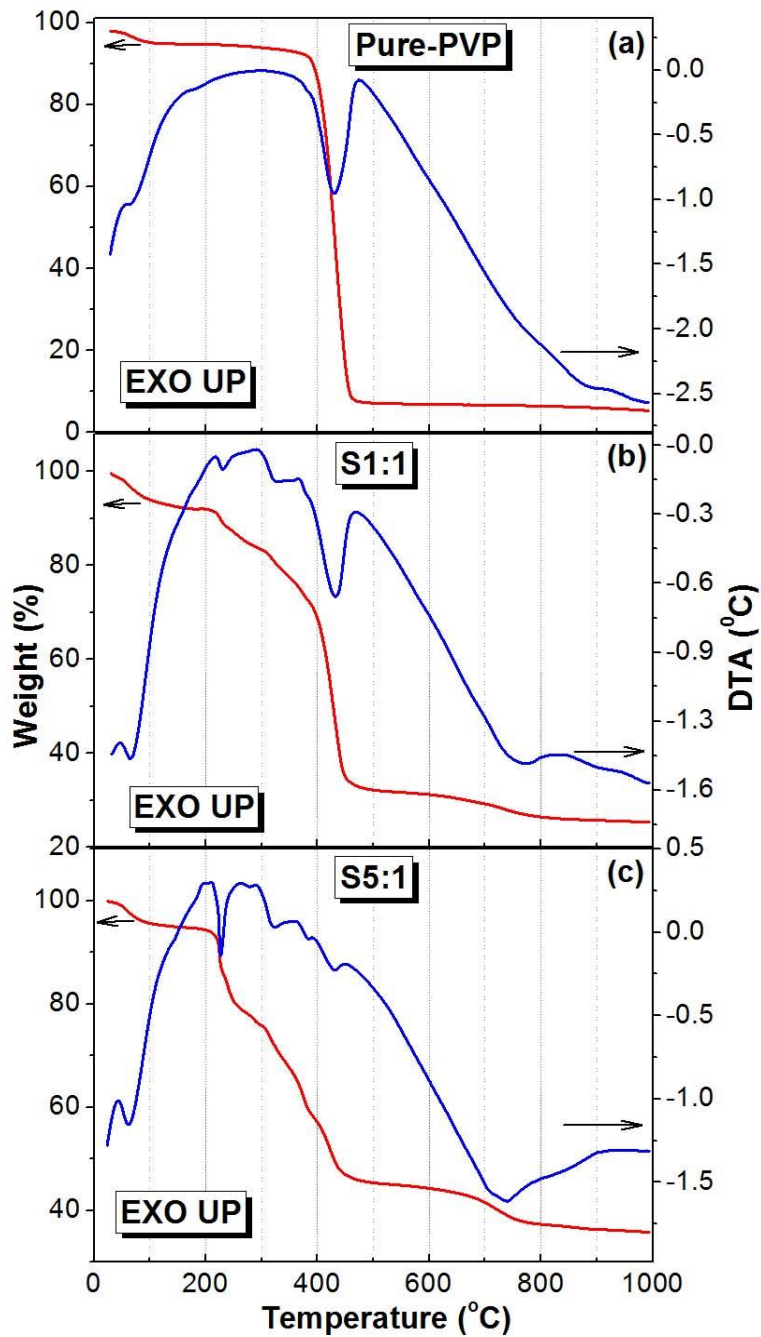

Figure 3: (Color online) Simultaneous TGA/DTA thermal analysis of the as produced samples. (a) Reference (commercial PVP), (b) S11, and (c) S51 samples. The endothermic peak of the reference sample at 430°C indicates PVP degradation. In panels (b) and (c), the peaks at ~800°C indicate the beginning of the crystallization of the YBCO phase.

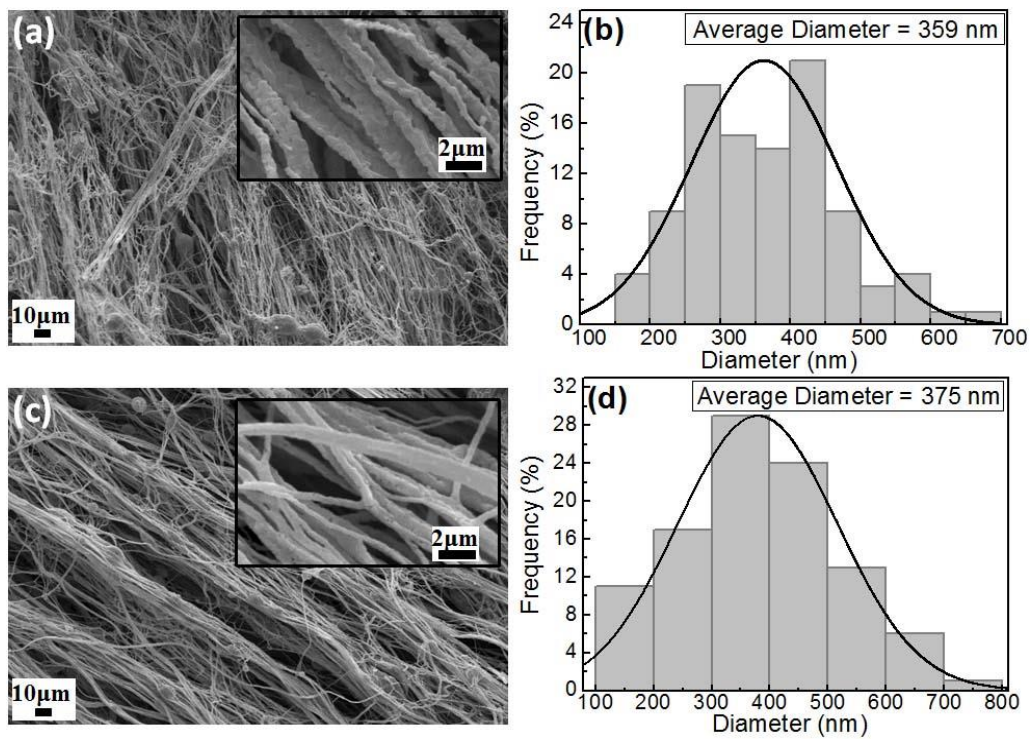

Figure 4: SEM micrographs and diameter distribution of the samples after all heat-treatments and oxygenation. (a) Micrograph of the S51 sample and (b) the diameter distribution of its nanofibers. (c) Micrograph of the S11 sample and (d) the diameter distribution of its nanofibers.

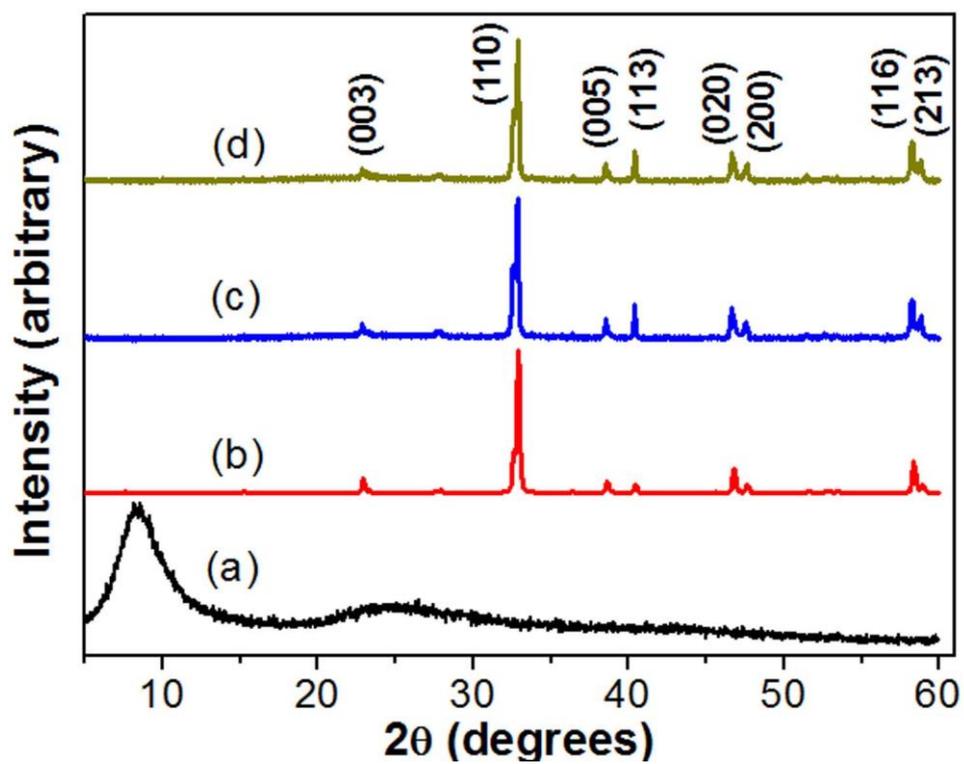

Figure 5: (Color online) XRD diffractograms of the samples. (a) A sample in the green state (S51), (b) reference YBCO pattern (JCPDS-78-2273), and (c) S51 and (d) S11 YBCO nanofibers after sintering.